\documentclass[journal,twoside,web]{ieeecolor}
\usepackage{tmi}
\usepackage{cite}
\usepackage{amsmath,amssymb,amsfonts}
\usepackage{algorithmic}
\usepackage{graphicx}
\usepackage{textcomp}

\usepackage{pifont}
\newcommand{\xmark}{\ding{55}}%
\newcommand{\cmark}{\ding{51}}%
\usepackage{multirow}
\usepackage{booktabs}
\usepackage{url}     
\usepackage{algorithm,algorithmic}
\usepackage{array}
\newcolumntype{P}[1]{>{\centering\arraybackslash}p{#1}}
\usepackage{stmaryrd}

\def\BibTeX{{\rm B\kern-.05em{\sc i\kern-.025em b}\kern-.08em
    T\kern-.1667em\lower.7ex\hbox{E}\kern-.125emX}}
\begin{document}
\bstctlcite{IEEEexample:BSTcontrol}

\title{Surface Vision Transformers: Flexible Attention-Based Modelling of Biomedical Surfaces}
\author{Simon Dahan, Hao Xu, Logan Z. J. Williams, Abdulah Fawaz, Chunhui Yang, Timothy S. Coalson, Michelle C. Williams, David E. Newby, A. David Edwards, Matthew F. Glasser, Alistair A. Young, Daniel Rueckert, \IEEEmembership{Fellow, IEEE}, and Emma C. Robinson 
\thanks{This work has been submitted to the IEEE for possible publication. Copyright may be transferred without notice, after which this version may no longer be accessible}
\thanks{S. Dahan, H.Xu, L.Z.J. Williams, A. Fawaz, A.D. Edwards, Alistair Young and E.C. Robinson are with the School of Biomedical Engineering and Imaging Sciences, King's College London, UK (e-mail:simon.dahan@kcl.ac.uk; emma.robinson@kcl.ac.uk). }
\thanks{M.F. Glasser, T.S. Coalson, C. Yang, are with the Departments of Radiology and Neuroscience, Washington University Medical School, USA.}
\thanks{D. Rueckert is with the
BioMedIA group, Department of Computing, Imperial College London,
UK and also with the Institute for AI and Informatics in Medicine,
Klinikum rechts der Isar, Technical University of Munich, Germany. }
\thanks{M.C. Williams and D.E. Newby are with the University/BHF Centre for Cardiovascular Science  University of Edinburgh, UK.}
}

\maketitle

\begin{abstract}

Recent state-of-the-art performances of Vision Transformers (ViT) in computer vision tasks demonstrate that a general-purpose architecture, which implements long-range self-attention, could replace the local feature learning operations of convolutional neural networks. In this paper, we extend ViTs to surfaces by reformulating the task of surface learning as a sequence-to-sequence learning problem, by proposing patching mechanisms for general surface meshes. Sequences of patches are then processed by a transformer encoder and used for classification or regression. We validate our method on a range of different biomedical surface domains and tasks:  brain age prediction in the developing Human Connectome Project (dHCP), fluid intelligence prediction in the Human Connectome Project (HCP), and coronary artery calcium score classification using surfaces from the Scottish Computed Tomography of the Heart (SCOT-HEART) dataset, and investigate the impact of pretraining and data augmentation on model performance. Results suggest that Surface Vision Transformers (SiT) demonstrate consistent improvement over geometric deep learning methods for brain age and fluid intelligence prediction and achieve comparable performance on calcium score classification to standard metrics used in clinical practice. Furthermore, analysis of transformer attention maps offers clear and individualised predictions of the features driving each task. Code is available on Github: \url{https://github.com/metrics-lab/surface-vision-transformers}.


\end{abstract}


\section{Introduction}
\label{sec:introduction}

The practice of modelling biomedical data as surfaces spans broad domains including cardiac \cite{J.Knuuti2019,C.Mauger2018,C.Mauger2019,H.Xu2021}, brain \cite{B.Fischl2008,M.Glasser2013,M.Glasser2016,E.Robinson2014,E.Robinson2018,R.Dimitrova2021,garcia2018dynamic}, respiratory \cite{nakao2021deformation} and musculoskeletal \cite{rajamani2007statistical,williams2010anatomically} imaging, with applications in 
biophysical and shape modelling \cite{cootes2001statistical,garcia2018dynamic,williams2010anatomically}, diagnostic stratification \cite{C.Mauger2019,wu2021cortical,zarei2013cortical,sugihara2017distinct,peng2015surface,raznahan2010cortical,hong2018multidimensional}, mapping of cortical organisation \cite{M.Glasser2016,R.Dimitrova2021,L.Williams2021}, 
and more. While the shapes of meshes may vary greatly, ultimately, all problems may be reduced to analysis of functions over tessellated, deformable meshes. Despite this, there is no unified geometric deep learning framework for studying all these problems.

Recently, Dosovitskiy et al. proposed the Vision Transformer, which sought to extend the use of self-attention transformer architectures, used in Natural Language Processing (NLP), to imaging data, by treating computer vision tasks as a sequence-to-sequence learning problem. In \cite{ViT}, RGB images were split into a grid of $16 \times 16 \times 3$ non-overlapping patches, forming a sequence that was passed to a vanilla transformer encoder \cite{A.Vaswani2017}
for image classification. Results showed that the ViT was able to offer a scalable solution that outperformed comparable CNNs when pre-trained on very large datasets. Lately, many improvements in the architecture have been proposed \cite{Ze.Liu2021}, leading to vision transformers now being considered as generic vision backbones for a broad range of tasks including image classification \cite{Ze.Liu2021,Z.Liu2022}, detection \cite{N.Carion2020} and segmentation \cite{S.Zheng2020}.

In this paper we propose a Surface Vision Transformer (\emph{SiT}), which extends the ViT to general surfaces through proposing mechanisms for surface patching. In particular, we extend from previous work \cite{S.Dahan2022}, in which the method was validated on neurodevelopmental phenotype prediction using cortical surface data from the Developing Human Connectome Project (dHCP), to also investigate fluid intelligence prediction using multimodal cortical imaging data from the Human Connectome Project (HCP) \cite{M.Glasser2013,M.Glasser2016}, and prediction of high coronary artery calcium score (a risk biomarker highly predictive of cardiovascular disease) using cardiac meshes from the Scottish Computed Tomography of the Heart (SCOT-HEART) trial \cite{DE.Newby2018}.

Surface deep learning architectures offer important opportunity for cortical applications since human brain morphology and functional topography varies considerably across individuals, in ways that violate the assumptions of traditional image registration \cite{M.Glasser2016}, and thereby limit the sensitivity of population-based comparisons. 
Since cortical surface representations encode more biologically meaningful geodesic distances between cortical areas  \cite{VanEssen2011,M.Glasser2013, E.Robinson2014, E.Robinson2018}, they are able to support more precise comparison of features of cortical micro and cyto-architecture \cite{R.Dimitrova2021,M.Glasser2016}, shape \cite{L.Williams2021} and function \cite{L.Williams2021,M.Glasser2016}. This is important since changes to cortical organisation are implicated in numerous neurological \cite{wu2021cortical,zarei2013cortical}, psychiatric \cite{sugihara2017distinct,peng2015surface} and developmental disorders \cite{raznahan2010cortical,hong2018multidimensional}.
Equally, surface-based cardiac models are essential for characterising the biomechanical properties of the atria and ventricles, for instance in patients with suspected coronary artery disease \cite{J.Knuuti2019}.

Recently, \cite{A.Fawaz2021} benchmarked a number of geometric deep learning (gDL) methods on cortical phenotype regression and segmentation, and found that these methods typically involve trade-offs between computational complexity, feature expressivity, and rotational equivariance of the models. The objective of this paper is to propose the SiT as a competitive alternative to these frameworks. The key contributions of this paper, relative to \cite{S.Dahan2022}, are as follows: 

\begin{itemize}
    \item  We demonstrate two general frameworks for sequence-to-sequence modelling of biomedical surfaces, which patch surfaces either via projection to a regularly tessellated icosphere, or through use of a finite element model \cite{C.Mauger2019}.
    \item Surface Vision Transformers (\emph{SiT}) are compared against geometric CNNs and traditional machine learning and demonstrate competitive results over a diverse range of cortical and cardiac prediction tasks.
    \item We extend beyond \cite{S.Dahan2022} to improve optimisation through data augmentation and show this leads to significant gains for the challenging task of birth age prediction.
    \item Maps of attention extracted from the \emph{SiT} encoder are displayed on the input space for the fluid intelligence task, to demonstrate that \emph{SiTs} can generate interpretable visualisations of the key features driving each task.   
\end{itemize}

\section{Related Works}

\begin{figure*}[!thb]
  \centering
\makebox[\linewidth]{
	\includegraphics[width=2.0\columnwidth]{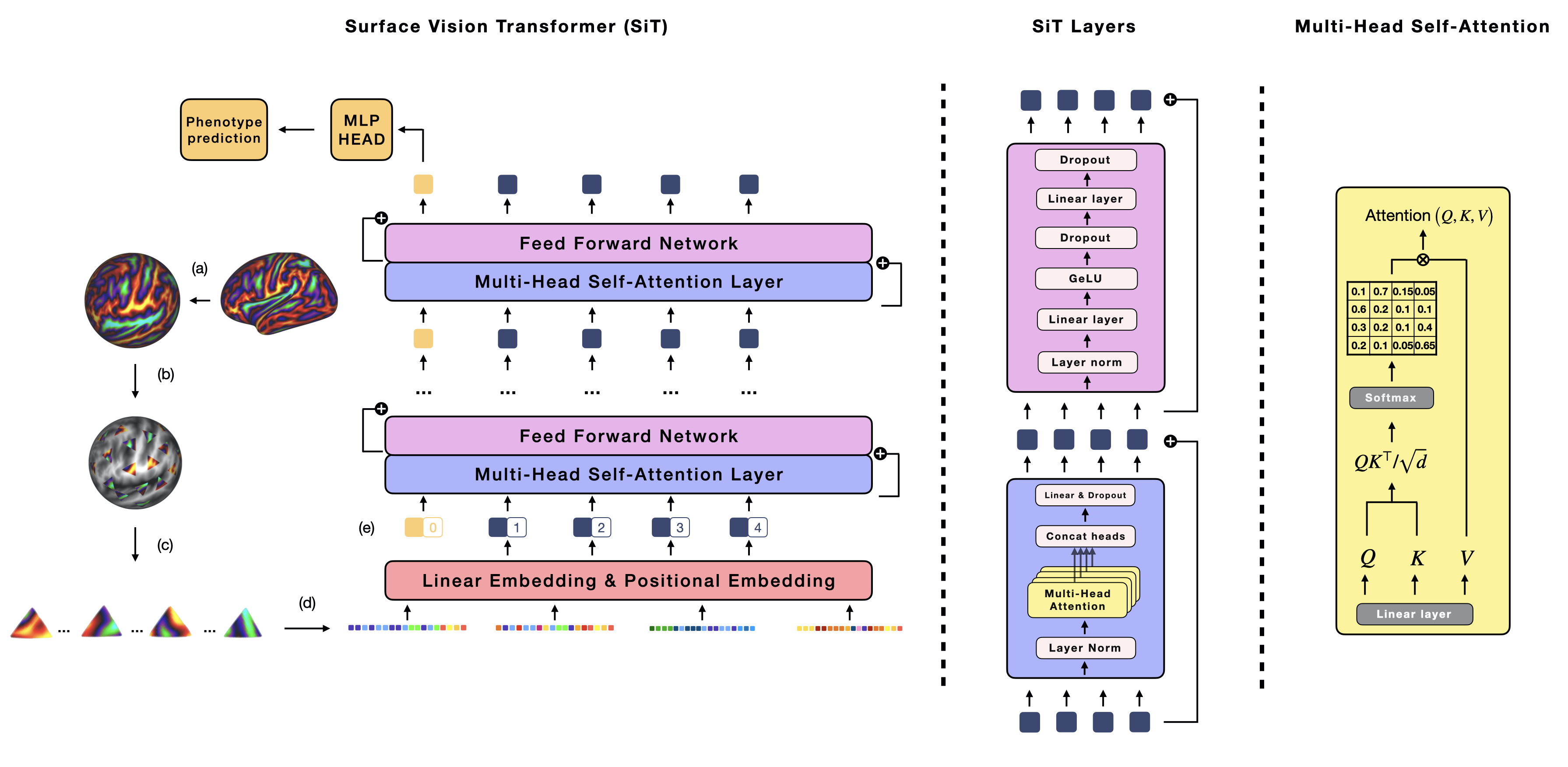}}
\caption{Surface Vision Transformer (\emph{SiT}) architecture. The cortical data is first resampled (a), using barycentric interpolation, from its template resolution (32492 vertices) to a sixth order icosphere (mesh of 40962 equally spaced vertices). The regular icosphere is divided into triangular patches of equal vertex count (b, c) that fully cover the sphere (not shown), which are flattened into feature vectors (d), and then fed into the transformer model. A positional embedding and an extra token for classification/regression is added to the sequence (e).}
\label{figure:sit}
\end{figure*}

\subsection{Geometric Deep Learning}
Geometric deep learning (gDL) is a field that seeks to translate concepts from Euclidean deep learning, to non-Euclidean manifolds such as graphs, surfaces or point clouds. Studying surfaces is particularly challenging as many gDL frameworks would in principle be suitable for studying mesh topologies, including: graph convolutional networks \cite{M.Defferrard2017}, which learn convolutions in the graph spectral domain, through polynomial approximation of the eigenvectors of the graph Laplacian; spherical spectral CNNs which fit convolutions using spherical harmonics or Wigner-D matrices \cite{T.Cohen2018}; spatial-templating approaches, which approximate Euclidean CNNs by fitting spatially localised filters to the surface \cite{F.Zhao2019,F.Monti2017}; and point cloud networks, which take coordinates as inputs and process data using multiple shared multi-layer perceptron units, followed by a permutation invariant function, to learn global feature vectors that can then be used for classification. Recent work \cite{A.Fawaz2021} showed that each architecture generates very different solutions to the same biomedical problem, with variable performance across different tasks. 

\subsection{Attention}

The concept of attention was introduced to the domain of Natural Language Processing (NLP), in the context of neural machine translation, to solve the bottleneck present in Recurrent Neural Networks (RNN) that limits the flow of contextual information in long sentences. Attention was first implemented in the decoder part of RNNs, by learning output tokens that weight the contribution of different words from the input sequence, in such a way to selectively focus on the most meaningful content \cite{D.Bahdanau2014, M.Luong2015}. Subsequently in \cite{A.Vaswani2017}, Vaswani et al introduced the self-attention (SA) mechanism of transformers, which further improved the modelling of long-range dependencies in sequences by relating all elements in each sequence via a pairwise alignment score (a scaled dot-product in \cite{A.Vaswani2017}). Despite the quadratic complexity of such an operation ($O(n^2)$ with n the sequence length), transformers have become the de-facto architecture in NLP leading to major breakthroughs in language understanding with the BERT model  \cite{J.Devlin2019} and GPT models \cite{A.Radford2018,A.Radford2019}, as notable examples.

In contrast, computer vision and medical imaging have mostly been pushed forward by developments in CNNs, where the locality and weight sharing properties of the convolution operation have created sample-efficient architectures that can generalise to a broad range of tasks. However, this inductive bias towards locality also has drawbacks, since it induces a limited receptive field that impairs the modelling of long-range spatial dependencies between distant parts of an image. This prevents CNNs from efficiently modelling processes that are diffuse in space and/or time; something that is known to be true for a wide range of biomedical applications, including multi-organ segmentation \cite{Y.Tang2021}, cognitive and neurodevelopmental modelling \cite{BH.Kim2021,R.Dimitrova2021} and estimation of the spatio-temporal dynamics of the heart \cite{D.Perperidis2005, C.Mauger2019}.

As a result, many attempts were made to introduce some add-on form of attention mechanisms into CNNs, as a way to incorporate more informative features across images. In one of the first examples, applied to video classification, \cite{Wang2018} proposed a non-local block that computed self-attention for CNN feature maps, by estimating the dot-product similarity of activations across all map locations, as a way to aggregate non-local information.
Limited by the quadratic cost of such operations at a pixel-level, these attention maps were only computed at low-resolution. 
Such non-local attention block modules have been inserted into the encoders of U-Net-like medical imaging segmentation architectures  
for vertebrae \cite{S.Joutard2019} and brain segmentation \cite{Z.Wang2018}. Similarly, attention-gate mechanisms have been developed, which identify the most salient regions in CNN feature maps, to increase their contributions to the learning process \cite{S.Jetley2018,F.Wang2017, J.Schlemper2018}; 
and Squeeze and Excitation blocks have been implemented for dynamic channel-wise feature recalibration, used for natural image classification \cite{J.Hu2017}, and adapted to improve 2D-slice brain segmentation in \cite{AG.Roy2019}.

Regardless, any use of attention blocks within CNNs remains inherently limited by the inductive biases that limit the learning of long-range associations. 
Therefore, in the Vision Transformer \cite{ViT} Dosovitskiy et al. proposed a paradigm shift, in which they suggested image recognition tasks should be reformulated as a sequence-to-sequence learning problem.
By doing so, they showed that general-purpose transformer architectures \cite{A.Vaswani2017} could be used for natural image classification; thereby demonstrating the benefits of using self-attention on image patches to improve modelling of global-context without relying on strong spatial priors. 
Since then, the performance of end-to-end Vision Transformers has challenged even highly optimised CNNs for image classification \cite{H.Touvron2020, Ze.Liu2021, X.Chen2021,Z.Liu2022}, object detection \cite{X.Zhu2020}, semantic segmentation \cite{S.Zheng2020} and video understanding \cite{A.Arnab2021}.

Similarly in medical imaging, recent years have seen emergence of many vision transformers models replacing CNNs. Such models have demonstrated 
enhanced global context, leading to both qualitative and quantitative improvements for histopathology \cite{C.Nguyen2021} and brain tumor segmentation \cite{Q.Jia2021, 
A.Hatamizadeh2022}, tumor classification \cite{Z.Shao2021} and polyp detection \cite{Z.Shen2021}. Pure (end-to-end) transformers have even shown advantages over CNNs in the low-data regime for brain segmentation \cite{D.Karimi2021}. While, development of transformers for 3D data is challenging due to the quadratic cost of self-attention, it may be mitigated by the use of hybrid architectures that combine transformer encoders with CNNs layers \cite{Q.Jia2021,Z.Shen2021,D.Kaizhong2021}, or through use of efficient transformer architectures with revised self-attention operations, used for medical imaging tasks such as \cite{ D.Kaizhong2021, A.Hatamizadeh2022}.

\section{Methods}

\subsection{Architecture}

The \emph{SiT} models translate surface understanding to a sequence-to-sequence learning task by reshaping the high-resolution grid of the input domain $X$,  into a sequence of $N$ flattened patches $ \widetilde{X} = \left [ \widetilde{X}^{(0)}_1, ..., \widetilde{X}^{(0)}_N \right] \in \mathbb{R}^{N\times (VC)}$ (V vertices, C channels). These are are first projected onto a $D-$dimensional sequence $X^{(0)} = \left [ X^{(0)}_1, ..., X^{(0)}_N \right]\in \mathbb{R}^{N\times D}$, using a trainable linear layer. Then, an extra $D$-dimensional token for regression or classification is concatenated ($X^{(0)}_0$), and a positional embedding ($E_{pos} \in \mathbb{R}^{(N+1)\times D}$) is added, such that the input sequence of the transformer becomes $ X^{(0)} = \left [ X^{(0)}_0, ..., X^{(0)}_N \right] + E_{pos}$ (see Fig \ref{figure:sit}(b-e)).  The positional embedding is added to encode spatial information about the sequence of patches. While $E_{pos}$ can be fixed or trainable, here it is implemented in the form of a 1D learnable weights, similar to the approach employed in \cite{ViT}.

The \emph{SiT} generates patches from any regularly tessellated reference grid that supports down-sampling. For the cortical applications, this is achieved by imposing a low-resolution triangulated grid, on the input mesh, using a regularly tessellated icosphere (Fig \ref{figure:sit}(b)). For cardiac applications, this is achieved by subdividing a control mesh twice using the Catmull-Clark algorithm (Fig \ref{figure:cardiac}). More details are provided in Section \ref{sec-patching}.

The architecture of the \emph{SiT} is illustrated in Figure \ref{figure:sit}. The SiT network is made of $L$ consecutive transformer encoder blocks of \textit{Multi-Head Self-Attention} (MHSA) and \textit{Feed Forward Network} (FFN) layers, with residual layers in-between:
\begin{equation} 
\begin{aligned}
Z^{(l)}  & = \textbf{\emph{MSHA}}(X^{(l)}) + X^{(l)}\\
 X^{(l+1)} & = \textbf{\emph{FFN}}(Z^{(l)}) + Z^{(l)}\\
 & = \left [ X^{(l+1)}_0, ..., X^{(l+1)}_N \right] \in \mathbb{R}^{(N+1)\times D}
\end{aligned}
\label{eq:transformer1}
\end{equation}
Following standard practice in transformers, \emph{LayerNorm} \cite{J.Ba2016} is used prior to each MSHA and FFN layer (omitted for clarity in Eq \ref{eq:transformer1}). 
In the last output sequence, the regression token ${X}^{(L)}_0$ is used as input to the final Multi-Layer Perceptron (MLP) for prediction. 

Following \cite{S.Dahan2022}, the proposed \emph{SiT} model builds upon two variants of the data efficient image transformer or \textit{DeiT} \cite{H.Touvron2020}: \emph{DeiT-Tiny}, \emph{DeiT-Small}, adapted into smaller versions from the vanilla Vision Transformer (ViT) \cite{ViT}.
A number of $L=12$ layers or transformer encoder blocks is used for all SiT versions; however, they differ in their number of heads, hidden size or embedding dimension $D$, and in the number of neurons (\emph{MLP size}) in the FFN. In Table \ref{tab:vit-size}, details about the architectures are provided. The number of parameters in Table \ref{tab:vit-size} corresponds to the configuration for the dHCP dataset, as the size of the first linear layer depends on the number of vertices and channels of the patches, and therefore the overall number of parameters. As a comparison, with the 115 channels for the HCP cortical surface a \emph{SiT-tiny} has 8.6M parameters.

\begin{table}[h]
  \centering
  \setlength{\tabcolsep}{5pt}
  \begin{tabular}{lccccc}
    \toprule
    \textbf{Models} & Layers & Heads & Hidden size \textit{D} & MLP size  & Params.\\
    \midrule
    SiT-Tiny & 12 & 3 & 192 & 768 & 5.5M \\
    SiT-Small & 12 & 6 & 384 & 1536 & 21.6M \\
    \bottomrule
  \end{tabular}
  \caption{Architectures}
  \label{tab:vit-size}
\end{table}

\subsection{Multi-Head Self Attention and Feed Forward Layers}

The mechanism of self-attention is based on the computation of attention weights/scores between tokens in a sequence, to capture the relative importance of patches. For all Multi-Head Self-Attention layers (MHSA) layers, patches in the sequence are linearly projected into a triplet, \textit{\textbf{Q}uery}, \textit{\textbf{K}ey}, and \textit{\textbf{V}alue} ($Q,K,V \in \mathbb{R}^{(N+1) \times D}$ ), such that: $Q=X^{(l)}W_Q$, $V=X^{(l)}W_V$, $K=X^{(l)}W_K$. Where, for each patch $i$ in the sequence,  self-attention weights ($\omega_{i}= \left [ \omega_{i,j}  \right ]_{j=0...N}$) are estimated from the inner product: $\omega_{i,j}=\langle\,q_i,k_j\rangle$, between the query ($q_i$) of patch $i$, and keys from all patches  ($k_j, \forall j \in  \llbracket 0,N\rrbracket$). After scaling, a softmax layer is applied along rows  to derive the self-attention weight matrix $A = \textrm{Softmax} \left ( QK^\top / \sqrt{D}\right ) \in \mathbb{R}^{(N+1) \times (N+1)}$, where an illustration of the MHSA layer is provided in Fig \ref{figure:sit}. Finally, the output sequence ${SA}^{(l)}\left( Q, K, V \right) \in \mathbb{R}^{(N+1) \times D}$ is obtained by weighting columns $V$ based on the self-attention weights. 

\begin{equation}
    \textrm{SA}^{(l)}\left( Q, K, V \right) = \textrm{Softmax} \left(\frac{QK^\top}{\sqrt{D}}\right)V
\end{equation}

The Multi-Head Self-Attention layers (MHSA) run multiple self-attention operations in parallel on subdivided part of the input embedded sequence in order to capture different interactions between tokens in the sequence. In practice, it means that the input sequence of dimension D is divided into sub-parts of dimension $D_{h} = D/h$ and each head processes its own triplet $(Q_{h},K_{h},V_{h})$, such that:
\begin{equation}
    \begin{aligned}
     \textrm{$SA_{h}^{(l)}$}\left( Q_{h}, K_{h}, V_{h} \right) & = \textrm{Softmax} \left(\frac{Q_{h}K_{h}^\top}{\sqrt{D_{h}}}\right)V_{h}
    \end{aligned}
\end{equation}

Outputs of the MHSA heads of dimension $(N+1) \times D_h$ are concatenated together along the channel dimension and then linearly projected to preserve the sequence dimension $D$, with a residual connection:
\begin{equation} 
\begin{aligned}
Z^{(l)} & = \left [ SA^{(l)}_0, ..., SA^{(l)}_H \right]W^{(l)} + X^{(l)}
\end{aligned}
\label{eq:transformer}
\end{equation}

Each MHSA layer is then followed by a Feed Forward Network, which consists of a succession of Layer norm, two linear layers with GeLU activation and dropout. The linear Layer expands the dimension of the sequence of patches to $4\times D$ then reduces it to $D$. 

\subsection{Surface Patching}
\label{sec-patching}

Surface patching is implemented differently for the cortical and cardiac domains. For the cortex, all imaging data were first projected to a sphere as part of the dHCP \cite{A.Makropoulos2018} and HCP \cite{M.Glasser2013} pipelines. Spherical data were then resampled onto a regular sixth-order icosahedron with 40,962 equally spaced vertices using barycentric interpolation, then split into triangular patches, where each patch corresponds to all data points within one face of a second-order icosphere (153 vertices per patch). The sequence is thus made of 320 non-overlapping patches sharing only common edges (Fig. \ref{figure:sit} (a-c)).

\begin{figure}[!t]
  \centering
\makebox[\linewidth]{
	\includegraphics[width=1.0\columnwidth]{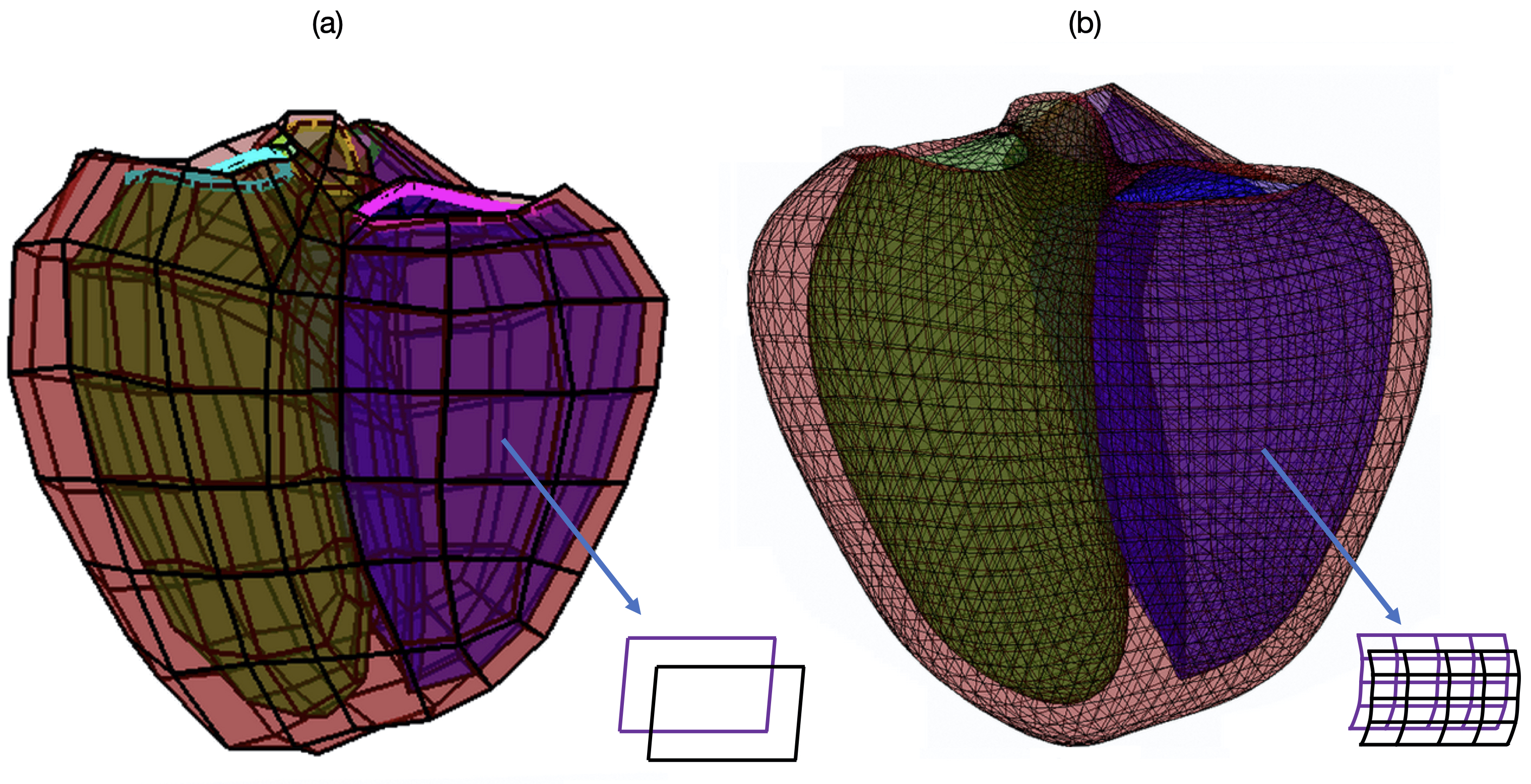}}
\caption{Biventricular coarse mesh (a), and final mesh (b), longitudinal view. The green, purple, and black surfaces are the LV endocardial, RV endocardial and biventricular epicardial surfaces.}
\label{figure:cardiac}
\end{figure}

By contrast the cardiac mesh is a non-closed surface, including inner (endocardial) and outer (epicardial) surfaces for the left and right ventricles. 
Illustration of the  mesh is provided in Fig. \ref{figure:cardiac}.  Coronary computed tomography angiography (CCTA) images part of the SCOT-HEART trial \cite{DE.Newby2018} were first segmented using a neural network algorithm \cite{H.Xu2021} and then biventricular meshes were fitted to the surface points extracted from the segmentation results by deforming a finite element model \cite{C.Mauger2019}. The biventricular subdivision surface model was initiated with a control mesh consisting of 388 vertices and 180 elements. A final mesh was obtained after subdividing the control mesh twice using the Catmull-Clark algorithm, giving $5\times 5=25$ vertices for each surface patch \cite{C.Mauger2018} as shown in Figure \ref{figure:cardiac}, redrawn from \cite{C.Mauger2018}. This resulted in a template mesh of 354 patches, where patches were subsequently merged in pairs to generate 177 larger patches 
(with 50 vertices per patch) which were then used for training.

\subsection{Optimisation}

To mitigate the lack of inductive biases in the architecture, transformers typically require large training datasets. 
In this paper, we therefore explore techniques for improving model generalisation: specifically pre-training and augmentation. 

\subsubsection{Pretraining} \label{sec-ssl} is performed through self-supervision, implemented as a \emph{masked patch prediction} (MPP) task, following the approach proposed in BERT \cite{J.Devlin2019}. This consists of corrupting some input patches at random, then training the network to learn how to reconstruct the full corrupted patches
. In this setting, we corrupt at random 50\% of the input patches, either replacing them with a learnable mask token (80\%), another patch embedding from the sequence at random (10\%) or keeping their original embeddings (10\%). To optimise the reconstruction, the mean square error (MSE) loss is computed only for the patches in the sequence that were masked. 
\subsubsection{Data augmentation} In extension to previous work \cite{S.Dahan2022} we additionally propose 
to augment the icosahedral patch selection by implementing $\pm \left \{5^{\circ},10^{\circ},15^{\circ},20^{\circ},25^{\circ},30^{\circ} \right \}$ rotations of the sphere around one of the x,y,z axis. These are implemented for HCP and dHCP experiments. Further optimisation strategies tailored for the specific datasets are presented in the corresponding sections. 

\subsection{Visualisation}
\label{section-visu}
Attention maps can be visualised on the input space by aggregating the attention weights across all transformer layers. To do so, here, the final attention weight matrix is $A^{(L)} = \textrm{Softmax} \left ( QK^\top / \sqrt{D}\right ) \in \mathbb{R}^{(N+1) \times (N+1)}$, where $Q=X^{(L-1)}W_Q$, $K=X^{(L-1)}W_K$ is recursively multiplied to the attention weight matrices of previous layers ($\left [ A^{(l)}  \right ]_{l=(L-1)...1}$). The identity is added to all attention matrices $A^{(l)}$ to take into account the contributions of the residual connections. The attention maps presented in Fig. \ref{figure:attention-map} were then obtained by extracting and normalising the first row of the resulting matrix, corresponding to the vector of attention scores between the classification/regression token and all other tokens in the sequence. This vector can be resolved at resolution ico2 and upsampled to resolution ico6 to support comparison to the original data.
As a different attention weight matrix is available per layer, but also per head $A^{(l)}_h$, we repeat this procedure to extract attention-maps for each head. As each transformer head being run in parallel, the self-attention operation should attend to different parts of the input sequence. 

\section{Results}

\begin{table*}[h]
  \footnotesize
  \centering
  \setlength{\tabcolsep}{10pt}
  \renewcommand{\arraystretch}{1.15}
  \begin{tabular}{l|c|c|c|c|c|c|c}
    \hline
    \multirow{2}{1.2cm}{\textbf{Methods}} &\multirow{2}{1.5cm}{\textbf{Pretraining}} &\multirow{2}{1.7cm}{\textbf{Augmentation}}  &\multicolumn{2}{c|}{\textbf{PMA}} & \multicolumn{2}{c|}{\textbf{GA - deconfounded}}  &\multirow{2}{1.0cm}{\textbf{Average}}\\
    \cline{4-7}
    &&&  \textbf{Template} & \textbf{Native}&   \textbf{Template} &\textbf{Native} \\
    \hline
    \hline
    S2CNN &\xmark & \cmark  & 0.63 $\pm0.02$& 0.73 $\pm0.25$  & 1.35 $\pm0.68$& 1.52 $\pm0.60$&1.06\\
    ChebNet &\xmark & \cmark  & 0.59 $\pm0.37$ & 0.77 $\pm0.49$& 1.57 $\pm0.15$& 1.70 $\pm0.36$& 1.16\\
    GConvNet &\xmark & \cmark   & 0.75 $\pm0.13$& 0.75 $\pm0.26$& 1.77 $\pm0.26$& 2.30 $\pm0.74$&1.39\\
    Spherical UNet & \xmark & \cmark  & 0.57 $\pm0.18$& 0.87 $\pm0.50$& \textbf{0.85} $\pm0.17$& 2.16 $\pm0.57$&1.11\\
    MoNet & \xmark & \cmark  & 0.57 $\pm0.02$ & \textbf{0.61} $\pm0.05$&   1.44 $\pm0.08$&1.58 $\pm0.06$&1.05\\
    \hline
    \hline
    SiT-tiny & \xmark & \xmark& 0.63 $\pm$0.01  & 0.77 $\pm$0.03 & 1.43 $\pm$0.01 & 1.75 $\pm$0.14& 1.15\\
    SiT-tiny & \xmark & \cmark& 0.69 $\pm$0.01 & 0.78 $\pm 5e^{-3}$ & 1.17 $\pm$0.06 & 1.36 $\pm$0.01&1.00\\
    SiT-tiny &  \cmark& \xmark& 0.58 $\pm$0.01 & 0.64 $\pm$0.06  & 1.40 $\pm$0.23 & 1.70 $\pm$0.10&1.08\\
    SiT-tiny &  \cmark& \cmark& 0.63 $\pm$0.01 & 0.67 $\pm$0.03 & 1.03 $\pm$0.09 & 1.31 $\pm$0.01&0.91\\
    \hline
    SiT-small &  \xmark& \xmark & 0.60 $\pm$0.02& 0.76 $\pm$0.03  & 1.14 $\pm$0.12 & 1.44 $\pm$0.03&0.99\\
    SiT-small &  \xmark& \cmark & 0.64 $\pm0.02$& 0.75 $\pm$0.01  & 1.14 $\pm$0.05& 1.22 $\pm$0.04&0.94\\
    SiT-small &  \cmark& \xmark& \textbf{0.55} $\pm$0.04& 0.63 $\pm$0.06 & 1.25 $\pm$0.06& \textbf{1.21} $\pm$0.22&0.91\\
    SiT-small &  \cmark& \cmark& 0.61 $\pm 1e^{-3}$& 0.71 $\pm$0.02& 1.02 $\pm$0.07& \textbf{1.21} $\pm$0.16&\textbf{0.89}\\
    \hline
    \hline
  \end{tabular}
  \caption{Results of SiT for the task of PMA and GA on Template and Native Space. Best MAE is reported and standard deviations for the three best models. Augmentations for gDL methods refer to non-linear warping, and to rotations for the SiTs.}  
  \label{tab:results-vit}
\end{table*}

We evaluate the performance of the SiT methodology on three tasks; 1) prediction of postmenstrual age at scan and gestational age at birth using neonatal data from the developing Human Connectome Project (dHCP) \cite{E.Hughes2017}, 2) prediction of fluid intelligence scores from the Human Connectome Project multimodal parcellation dataset (HCP)  \cite{M.Glasser2013,M.Glasser2016}, and 3) classification of cardiac surfaces between high and low coronary
artery calcium score (CACS) from the Scottish Computed Tomography of the Heart (SCOT-HEART) dataset \cite{DE.Newby2018}. All experiments were run using an NVIDIA RTX 3090 24GB GPU.

\subsection{dHCP experiments: Scan age and Birth age predictions}
\label{sec-dhcp}

\subsubsection{dHCP Dataset}

Data for this experiment corresponds to cortical surface meshes and metrics, derived from the third release of the developing Human Connectome Project (dHCP) \cite{E.Hughes2017}. Surfaces were extracted from T2- and T1-weighted Magnetic Resonance Imaging (MRI) scans using the dHCP pipeline \cite{A.Makropoulos2018, MKuklisova-Murgasova2012, A.Schuh2017,E.Hughes2017, L.Cordero-Grande2018}. 
Briefly, T2w and T1w scans were motion and bias corrected, brain-extracted, and segmented, using the the Draw-EM \cite{A.Makropoulos2014} algorithm, which parcellates scans into 9 tissue types: including cortical grey matter and 3 categories of white matter. White matter masks were then fused, a topology correction step was implemented and surfaces were generated through an iterative process of inflation and smoothing. Among the various surface-based features generated by the pipeline, four cortical surfaces metrics were used in this work: sulcal depth, curvature, cortical thickness and T1w/T2w myelination. Data were registered using the Multimodal Surface Matching algortihm \cite{E.Robinson2014,E.Robinson2018} to the left-right symmetric 40-week sulcal depth template from the dHCP spatiotemporal cortical atlas \cite{J.Bozek2018,L.Williams2021}. Subsequently, surfaces were then resampled to a sixth-order icosahedral mesh of 40,962 equally spaced vertices. Experiments were run on both \emph{template}-aligned data and unregistered (\emph{native}) data, and train/test/validation splits parallel those used in \cite{A.Fawaz2021}. 

A total of 588 images were utilised, acquired from term (born $\ge37$ weeks gestational age, GA) and preterm (born $<37$ weeks GA) neonatal subjects, scanned between 24 and 45 weeks postmenstrual age (PMA). Some of the preterm neonates were scanned twice: once after birth and again around term-equivalent age. The proposed framework was benchmarked 
on two phenotype regression tasks:  prediction of postmenstrual age (PMA) at scan, and gestational age (GA) at birth, where since the objective was to model PMA and GA as markers of healthy development, all preterms' second scans were excluded from the PMA prediction task, and all first scans were excluded from the GA regression task. This resulted in 530 neonatal subjects for the PMA prediction task (419 term/111 preterm), and 514 neonatal subjects (419 term/95 preterm) for the GA prediction task.

\subsubsection{Training}

The task of GA prediction is arguably more complicated than the PMA task, as it is run on scans acquired around term-equivalent age ($37-45$ weeks PMA) for both term and preterm neonates, and therefore is highly correlated to PMA at scan. Previous work by \cite{A.Fawaz2021} have shown the benefit of deconfounding the scan age for the task of birth age prediction, where for all gDL methods an additional 1D convolution was used to incorporate scan age as a confound, before the last fully connected layer used to make the birth age prediction. Here, a deconfounding strategy was employed where the scan age information was incorporated into the patch sequence by adding an extra embedding to all patches in the sequence before the transformer encoder. This was implemented using a fully connected network to project scan age to a vector embedding of dimension $D$ after batch-normalisation \cite{S.Ioffe2015}. 

The dHCP dataset is heavily unbalanced with more term babies than preterm babies. In extension to previous work, this class imbalance was addressed by adapting sampling during training. Subjects were split into 3 categories, which reflect the clinical subcategories of preterm birth \cite{spong2013defining}: over 37 weeks, between 32 and 37 and below 32 weeks. The original ratio of examples in each of these three categories was 1/7/11.

In all instances, the proposed transformer networks were compared against the best performing surface CNNs reported in \cite{A.Fawaz2021}: Spherical U-Net \cite{F.Zhao2019}, MoNet \cite{F.Monti2017}, GConvNet \cite{T.Kipf2017}, ChebNet \cite{M.Defferrard2017} and S2CNN \cite{T.Cohen2018}. Since the best performance in \cite{A.Fawaz2021}, and presented in Table \ref{tab:results-vit},  was achieved using data augmentation (with non-linear warped surface meshes augmentation), we extend from \cite{S.Dahan2022} to further investigate whether data augmentation would improve generalisation. We also investigated the stability of the model by reporting the standard deviation over 3 runs. Best performances were obtained by using SGD optimiser for all PMA tasks and for the task of GA with augmentation and Adam for the GA task without augmentation. A batch size of 256 was used for \emph{SiT-tiny} and 128 for \emph{SiT-small}.

\subsubsection{dHCP Results}
Results for the tasks of PMA at scan and GA at birth are reported in Table \ref{tab:results-vit}. The results for SiT-tiny and SiT-small are reported under a range of training conditions: 1) following training from scratch or using pretraining; 2) with and without augmentation (in this case adaptive sampling and rotations). All \emph{SiT} models were trained from scratch for 2,000 iterations and following pre-training for 1,000 iterations. 

Overall, \emph{SiT-tiny} and \emph{SiT-small} configurations consistently outperformed two of the gDL methods (GConvNet and ChebNet), with \emph{SiT-small} returning the best performances overall for the tasks of PMA-template and GA-native. On average for both tasks, 6/8 SiT training configurations achieved prediction errors less than 1.00 MAE and outperformed all the gDL methods, with the best averaged performance for SiT-small pretrained and augmented with 0.89 MAE compared to the best gDL model, MoNet with 1.05 MAE.  

For the task of PMA, the best performing SiT configurations were obtained with SiT-small pretrained, for both template and native configurations: 0.55/0.63. These results are competitive with MoNet: 0.57/0.61 (the best performing gDL network). The use of data augmentation did not improve the performance of SiT models for this task but seems to reduce the variance across runs, especially for the native configuration. 

The gain in performance following data augmentation is particularly noteworthy for the task of birth age prediction (GA-deconfounded). For all SiT configurations (except SiT-small trained from scratch) the use of data augmentation greatly boosted performance, while reducing the variance, returning an average of 0.29 MAE improvements across all four SiT augmented configurations. With data from native space, augmentation consistently improved results (by at least by 0.16 MAE) relative to the best performing gDL model for this sub-task, S2CNN with 1.52. The best performances were obtained for rotations in the range $\pm \left \{5^{\circ},10^{\circ}\right \}$. Bigger rotations appeared to limit the gain in performance but would probably have benefited from training for more iterations. 

Finally, across all tasks, performances of SiT models are much more consistent. For example, relative to Spherical U-Net, which is the best performing network on aligned surfaces, SiT performs much better on native data, dropping only from 1.02 (for template PMA) to 1.21 (for native PMA) whereas Spherical U-Net drops from 0.85 to 2.16 MAE. The drop in performance on this task is known to be related to Spherical U-Net's lack of rotational equivariance \cite{A.Fawaz2021}. Likewise between the tasks of PMA and GA prediction, the SiT shows a much lower drop in performance for the GA task than MoNet - which although rotationally equivariant (and therefore consistent on native and template domains) learns less expressive convolutional filters, parameterised as a mixture of Gaussians.

\begin{figure*}[h]
  \centering
\makebox[\linewidth]{
	\includegraphics[width=2.0\columnwidth]{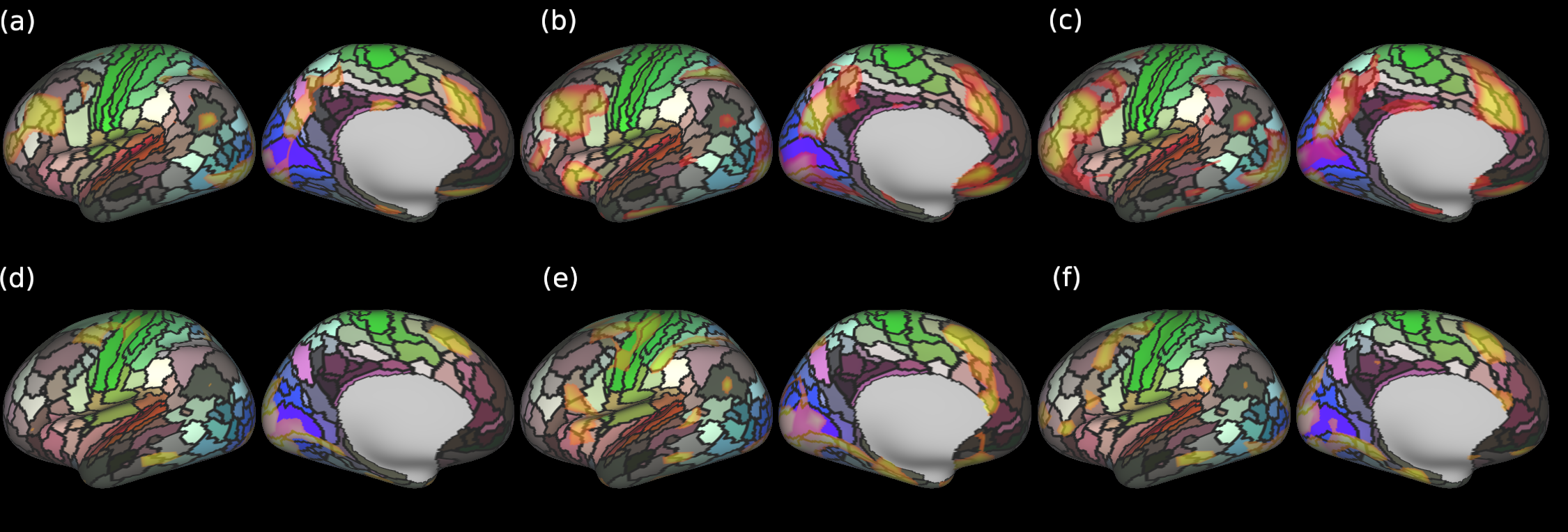}}
\caption{Attention maps overlaid on individual multi-modal parcellation for test subjects with high (first row) and low (second row) fluid intelligence scores 
Three attention heads are visualised per subject: (a) and (d) for head 1; (b) and (e) for head 2; (c) and (f) for head 3. Multimodal parcellations and attention maps are thresholded at 0.5 (range 0-1) and viewed on a very inflated left hemisphere view. 
}
\label{figure:attention-map}
\end{figure*}

\subsection{HCP experiments: Fluid intelligence predictions}




\subsubsection{Dataset} Data from 446 healthy individuals scanned as part of the HCP were used. Acquisition and minimal preprocessing pipelines are described in \cite{M.Glasser2013}. Cortical metrics correspond to the 115 features used for the HCP multimodal parcellation (MMP) \cite{M.Glasser2016} and include cortical thickness, T1w/T2w myelin, curvature, as well as 20 task fMRI maps and 77 resting state fMRI maps (derived from group ICA followed by weighted dual regression), a mean task-fMRI activation map and 9 visuotopic maps derived from weighted regression of hand-engineered retinotopic spatial maps, and 5 artefact maps. For more details on feature generation and pre-processing see \cite{M.Glasser2016}.

\subsubsection{Training}

The performance of the SiT was benchmarked for the task of fluid intelligence prediction - evaluated from the number of correct responses to the Penn progressive matrices (PMAT) task. To prevent mixing of data across families, examples were split into five folds, keeping all members of a family within a single fold. 

Best performance was achieved through implementing 
strong regularisation to prevent overfitting. A dropout of 0.5 was used before the first linear embedding and dropout of 0.3 was used in the FFN. For each split, models were trained for 200 iterations, as more iterations lead to overfitting, using a batch size of 128. Best correlation scores were obtained with the use of an SGD optimiser and a learning rate of $5e^{-4}$.  Data were augmented with rotations of the cortical surface and adaptive sampling, specifically by oversampling the subjects with low fluid intelligence scores, who were under-represented in the dataset. However, such augmentation did not strongly improve performance for this task.

Since fluid intelligence is most commonly estimated from functional brain connectivity, the method was compared against a spatio-temporal graph convolution neural network MS-G3D \cite{Z.Liu2020}  which 
was adapted to the analysis of cortical connectomes in \cite{S.Dahan2021}.
Functional data for this model comes from the first session (15mn - 1200 frames - 0.72s/frame) of the Human Connectome Project (HCP) S1200 release where individual-subject node timeseries were derived following group-wise independent component analysis (Group-ICA) and dual-regression \cite{C.Beckmann2009,L.Nickerson2017} to spatial maps (functional nodes, 15 to 300) and associated time courses. The MS-G3D was trained on functional connectivity matrices, derived from the correlation of ICA nodes' timeseries, where the results reported here deviate from what was reported in \cite{S.Dahan2021}, since this paper uses a only subset of the full HCP dataset (446/1003) corresponding to the datasets for which HCP multimodal parcellations are also available. Models were trained for 2000 iterations, and optimised with Adam and a learning rate of $3e^{-4}$, with a batch size of 128. Best performance was achieved for an ICA dimensionality of 100.

\subsubsection{Results}

\begin{table}[h]
\centering
\setlength{\tabcolsep}{6pt}
\begin{tabular}{lc}
        \toprule
        \textbf{Methods} & \textbf{Correlation} \\
        \toprule
        MS-G3D-ICA100 \cite{S.Dahan2021}  & 0.26 $\pm$ 0.08 \\
        \midrule
        SiT-tiny & 0.35 $\pm$ 0.03 \\
        \bottomrule
        \\
\end{tabular}
\caption{Fluid Intelligence results. Mean Pearson correlation scores are reported with standard deviation across 5-folds.}
\centering
\label{Table:fluid_intelligence_results}
\end{table}

The task of fluid intelligence prediction from high resolution cortical metrics is highly challenging as models must relate complex sources of information that are distributed across the whole brain. This need for long-range context should therefore benefit from self-attention.  

Results in Table \ref{Table:fluid_intelligence_results}, demonstrate that the SiT outperformed the spatio-temporal graph network MS-G3D, on the subset of 446 subjects used for this experiment. This score of 0.35 also outperforms the results for MS-G3D in \cite{S.Dahan2021} that report mean correlation of 0.325 across 5 folds, when trained on the full HCP dataset (1003), and compare strongly against other results in the literature \cite{J.Rasero2020,U.Pervaiz2020}. 
The attention maps, shown in Figure \ref{figure:attention-map}, visualise the regions most attended to during prediction for two individuals that each had
high and low fluid intelligence scores respectively, and were correctly predicted by the model. In both cases, attention focuses on association areas in the frontal and parietal lobes. This aligns with theory since these areas reflect regions most evolved relative to non-human primates. The combined attention to frontal and parietal regions, observed particularly for the subject with high intelligence, reflects  previous studies \cite{I.Deary2010} which suggests interactions between these spatially-distant regions are important for working memory.

. 

\subsection{SCOT-HEART experiments: high calcium classification}

\subsubsection{Dataset}
\label{section-cardiac-data}

Data for this experiment consisted of biventricular surface meshes, comprising endocardial and epicardial surfaces of the left ventricle (LV) and right ventricle (RV), customised to patients with suspected coronary artery disease, who participated in the Scottish Computed Tomography of the Heart (SCOT-HEART) trial \cite{DE.Newby2018}. We used coronary artery calcium score (CACS) as a surrogate biomarker, since it is known to be highly predictive of adverse cardiovascular events and associated with changes in heart geometry (mass and volume)\cite{M.Budoff2018,H.Bakhshi2017}. Relationships between heart geometry and cardiovascular risk factors inform mechanisms of heart disease and suggest treatment plans \cite{C.Mauger2019}. A healthy sub-cohort was identified as 248 participants with CASC $<300$ AU, body-mass index $<30$ kg/m2, no evidence of obstructive coronary artery disease on CCTA, and no documented hypertension, smoking history or diabetes mellitus. A high CACS sub-cohort was identified as CACS $>300$ AU with 367 participants. Figure \ref{figure:cardiac} (b) shows an element of the coarse mesh and for each element endocardial and epicardial surfaces there are $25\times 2=50$ corresponding vertices from the final mesh.

\subsubsection{Training \& Results} The performance of the SiT was benchmarked for the task of cardiac mesh classification between healthy and high CASC against a baseline logistic regression model, using standard LV and RV volumes and LV mass as predictors, with five-fold cross-validation. A mean cardiac shape was constructed using Procrustes alignment and the displacement of the biventricular meshes from the mean shape ($dx, dy, dz$) and the wall thickness ($t$) forms the 4 feature channels used for the \emph{SiT} models. Models were trained using five-fold cross validation and best performance was achieved using Adam optimiser and a learning rate of $3e^{-4}$. 


\begin{table}[h]
\centering
\setlength{\tabcolsep}{10pt}
\begin{tabular}{lc}
        \toprule
         \textbf{Methods} &  \textbf{AUC} \\
        \midrule
        Logistic regression & 0.750 \\
        \midrule
        SiT-tiny &  0.738 $\pm$ 0.04 \\
        SiT-small& 0.742 $\pm$ 0.03\\
        \bottomrule
        
        \\
\end{tabular}
\caption{Cardiac mesh classification results. SiT-tin was trained following a five folds cross-validation, and compared against a baseline logistic regression.}
\centering
\label{Table:cardiac_results}
\end{table}

Best performances (reported in Table \ref{Table:cardiac_results}) were obtained using the larger patches of 50 vertices. SiT models with smaller patches overfit dramatically, and would probably benefit from specific data augmentation and adaption of the embedding dimension to account for the small number of vertices in the patch sequence. \emph{SiT} models obtained good performances using mesh displacement and wall thickness, similar to the ones obtained with logistic regression used as standard clinical metrics. 

\section{Discussion}

In this paper, we build from \cite{S.Dahan2022}, to show that SiTs can improve over convolutional gDL frameworks, in terms of both performance and interpretability. The potential of the proposed framework was validated across a range of domains and tasks.

Results from experiments on dHCP phenotype regression, show that the SiT performs competitively with the best performing gDL networks, on data that has been pre-aligned, and shows far less drop in performance on unregistered data, relative to the best performing gDL network (Spherical U-Net \cite{F.Zhao2019}). This suggests that SiT is able to encode a degree of transformation invariance, which was further enhanced (relative to \cite{S.Dahan2022}) by training with augmentations. As a result, \emph{SiT-small} returned the best performance of all networks on the challenging GA prediction task and lowest error across all tasks. Likewise, for HCP fluid intelligence regression, the SiT trained on cortical imaging data strongly outperformed a graph convolutional network trained only on functional connectomes. 

At the same time, when validated on the fluid intelligence task, the SiT returned highly interpretable individualised attention maps, which highlighted regions already known to be highly important for working memory \cite{I.Deary2010}. These results parallel the visualisation of PMA prediction in \cite{S.Dahan2022}, which demonstrated the model was attending to a diffuse range of brain regions known to be associated with early neonatal cortical maturation.


Beyond cortical surface analysis, this paper extended the icosahedral patching methodology of \cite{S.Dahan2022}, to propose an alternative framework for patching non-closed surfaces that leverage finite element models. This approach was validated for cardiac mesh models by classifying calcium levels (as a biomarker of cardiovascular risk) from mesh features associated with cardiac shape and wall thickness.  
Results showed that the relationship with cardiac shape captured by SiT was similar to standard measures of cardiac mass and volume used in standard clinical practice. Nevertheless, future work would look to extend the features utilised by the SiT, for example to include measures of wall curvature, biomechanical measures of wall strain, or explore motion in space and time, to support computation of a risk score with improved prognostic significance. 

While performance gains were shown for the \emph{SiT} through pre-training \cite{ViT,H.Touvron2020}, regularising and incorporating data augmentation, the model still overfitted on the relatively small datasets of the HCP fluid intelligence task and the SCOT-HEART trial. It is possible that the domain gap between natural and medical imaging impacts the potential gains of pre-training strategies. Specific schemes for pre-training medical transformers and adapted losses have been shown to improve performance\cite{Y.Tang2021,F.Shamshad2022}, 
specifically for 3D medical segmentation, where tailored proxy tasks of self-supervision improved the learning of the underlying pattern of human anatomy, and achieved state-of-the-art results on multi-organ segmentation datasets \cite{Y.Tang2021}. 

Finally, while the presented results clearly emphasise the advantages of  
modelling long-range dependencies for biomedical surface applications, 
in some cases, the lack of inductive biases of transformers have been shown to impair their learning 
\cite{X.Chen2021}. Lately, therefore some studies have been reintroducing forms of inductive biases in end-to-end vision transformer architectures, reusing some concept of locality from convolutions. With Hierarchical Transformers, Liu et al \cite{Ze.Liu2021} introduced local forms of attention with sliding windows and improved on the limitations of the vanilla ViT for more complex tasks such as detection or semantic segmentation \cite{Ze.Liu2021, Z.Liu2022}. 
Such hierarchical approaches, like the Swin Transformer \cite{Ze.Liu2021}, have also been used to surpass the ViT for 3D brain segmentation tasks \cite{A.Hatamizadeh2022}. 

\bibliographystyle{IEEEtran}
\bibliography{main}

\end{document}